\documentstyle[12pt,preprint,eqsecnum,aps,prd,amsfonts]{revtex}

\begin{document}

\draft
\preprint{MADPH-96-934 }

\title{ On quark confinement dynamics}

\author{Theodore J. Allen }
\address{Department of Physics, University of Wisconsin, Madison,
	    WI 53706}
\author{M. G. Olsson}
\address{Department of Physics, University of Wisconsin, Madison,
	    WI 53706}
\author{Sini\v{s}a Veseli}
\address{Department of Physics, University of Wisconsin, Madison,
	    WI 53706}
\author{Ken Williams}
\address{Continuous Electron Beam Accelerator Facility \\
	Newport News, VA 29606, USA \\
	and \\
Physics Department, Hampton University, Hampton, VA 29668}

\date{\today}

\maketitle

\begin{abstract}
Starting from Buchm\"uller's observation that a chromoelectric flux tube
meson will exhibit only the Thomas type spin-orbit interaction, we show
that a model built upon the related assumption that a quark feels only a
constant radial chromoelectric field in its rest frame implies a complete
relativistic effective Hamiltonian that can be written explicitly in terms
of quark canonical variables.  The model yields linear Regge trajectories
and exhibits some similarities to scalar confinement, but with the advantage
of being more closely linked to QCD.

\end{abstract}


\newpage
\section{Introduction}
\label{intro}
The nature of quark confinement is an old but still active problem. The
transition to a linearly rising potential at large quark-antiquark
separation is evident from lattice simulations. The associated ``flux
tube'' field distribution has long been an attractive starting point to
develop classical and semi-classical meson models \cite{chodos} and
non-relativistic hybrid and glueball models \cite{isgur} as well as
relativistic quantized models of meson dynamics \cite{lacourse}.
Buchm\"uller \cite{buch} has proposed that a flux tube which is purely
chromoelectric in its rest (co-moving) frame must yield a spin-orbit
interaction that is only of the kinematic (Thomas) type. The result follows
classically, as we review in the next section, from the given fact that
there is no magnetic field in the co-moving frame.

In the present paper we show that this simple picture is far more powerful
than originally thought. We show that when one asks what fields in the
stationary laboratory frame are equivalent to those of a co-moving
chromoelectric flux tube, one is led to a complete relativistic effective
theory. Section \ref{classical} contains a relativistic version of
Buchm\"uller's argument based on the Thomas equation. The complete
effective Hamiltonian in the Bethe-Salpeter form is given in Section
\ref{effh}, and the reduction and consequent relativistic corrections are
worked out in Section \ref{corrections}.  In this section we also compare
our result with the relativistic corrections expected from scalar and time
component vector confinement as well as those expected from QCD.  The Regge
structure of our model is discussed in Section \ref{regge}.  Our
conclusions are summarized in Section \ref{con}.

\section{Buchm\"uller's picture of the spin-orbit interaction in QCD}
\label{classical}

The classical covariant treatment of a spinning charge moving in electric
and magnetic fields was first introduced by Thomas \cite{thomas} in
1927. This formalism automatically includes the ``Thomas precession''
effect.  The interaction energy of the spin magnetic moment is
\cite{jackson}
\begin{equation}
H_{\rm so} = -\frac{1}{m} {\bf s} \cdot {\bf G},
\label{so}
\end{equation}
where $m$ is mass of the particle and
\begin{equation}
{\bf G} = (\frac{g}{2}-1+\frac{1}{\gamma}){\bf B} - (\frac{g}{2} -
\frac{\gamma}{\gamma+1}) {\bf v} \times {\bf E} .
\label{gdef}
\end{equation}
Here, and in what follows, the particle charge has been absorbed into the
fields ({\it e.g.\/} $e{\bf B} \rightarrow {\bf B}$), and for simplicity we
assume ${\bf v} \cdot {\bf B} = 0$, which always will be true in the case
we consider. The gyromagnetic ratio $g$ is expected to take the value
$g=2$, but we consider the general case.

The above result can be re-expressed in terms of the quark co-moving
fields ${\bf B}'$ and ${\bf E}'$ by the usual Lorentz transformation
\begin{eqnarray}
{\bf B}_\parallel &=& {\bf B}'_\parallel,\\
{\bf E}_\parallel &=& {\bf E}'_\parallel,\\
{\bf B}_\perp &=& \gamma({\bf B}'_\perp + {\bf v} \times {\bf
E}'),
\label{blt}\\
{\bf E}_\perp &=& \gamma({\bf E}'_\perp - {\bf v} \times {\bf B}').
\label{elt}
\end{eqnarray}
Substituting the transformed fields into (\ref{gdef}), using ${\bf v} \cdot
{\bf B} = {\bf 0}$ and $\gamma^{-2} = 1 - v^{2}$, we obtain
\begin{equation}
{\bf G} = \frac{g}{2 \gamma} {\bf B}' + \frac{1}{\gamma + 1}{\bf v} \times
{\bf E}' .
\label{glt}
\end{equation}

Buchm\"uller's observation \cite{buch} follows directly from
(\ref{glt}). If the flux tube is pure chromoelectric in its rest
(co-moving) frame, then ${\bf B}'={\bf 0}$, and the spin-orbit term energy
is by (\ref{so}) and (\ref{glt})
\begin{equation}
H_{\rm so}
= -\frac{1}{m(\gamma+1)} {\bf s} \cdot ({\bf v} \times {\bf E}') .
\end{equation}
For low velocity quarks we have $\gamma \simeq 1$, ${\bf L} \simeq m {\bf
r} \times {\bf v}$. With the assumption ${\bf E}' = - a \hat{\bf r}' $,
where $a$ is the magnitude of the electric field, (sometimes called the
string tension in flux tube models), the spin orbit energy becomes
\begin{equation}
H_{\rm so} \simeq - \frac{a}{2 m^{2} r} {\bf s} \cdot {\bf L} .
\label{hsoth}
\end{equation}
This pure ``Thomas'' spin-orbit term arises in a natural way in the
Buchm\"uller picture. In QCD confinement the moving quark does not ``cut''
electric field lines as it moves since the electric field is carried along
in a roughly straight flux tube connecting the quarks.

For completeness we also mention the spin-orbit terms arising from other
assumptions often made about the confinement potential.  With a Lorentz
scalar confinement there is no magnetic field in any frame, so only the
purely kinematic Thomas term is present and the spin-orbit term
(\ref{hsoth}) again results. For the time component vector confinement
${\bf B} = {\bf 0}$ in the lab frame, and ${\bf E}' = -a\hat{\bf r} \simeq
{\bf E}$. The Lorentz transformations (\ref{blt}) then yield ${\bf B}' = -
\gamma {\bf v} \times {\bf E}'$.  By (\ref{so}) and (\ref{glt}) we then
find for a low particle velocity
\begin{equation}
H_{\rm so}^{\rm vector} \simeq \frac{a}{2 m^{2} r}(g-1) {\bf s} \cdot {\bf
L} .
\end{equation}
For $g=2$, the above equation becomes
\begin{equation}
H_{\rm so}^{\rm vector} \simeq \frac{a}{2 m^{2} r} {\bf s} \cdot {\bf L} ,
\end{equation}
which is of the same magnitude but opposite sign from the pure Thomas term.

Discussing QCD in an electrodynamic context is not unreasonable in this
model of quark dynamics given the assumptions we make about the physical
fields.  The particular flux tube field configuration present depends
crucially upon the nonabelian nature of QCD while the interaction of the
fields with the quarks does not in this case.  This is so because we assume
that the fields agree with that of a flux tube, wherein only one component
of the QCD field strength $G_{\mu\nu}$, namely the radial chromoelectric
field, is nonzero in a co-moving frame.  Thus we may write the $SU(3)$
field strength as ${G}_{\mu'\nu'}= G^a_{\mu'\nu'} T^a = { E^a
T^a}(\delta_{\mu' 0'}\delta_{\nu' r'} - \delta_{\mu' r'}\delta_{\nu' 0'})$
in a frame that is moving with the same velocity as the quark. As the field
tensor has only one non-zero component in this frame, we find that the
Lorentz components of the QCD field strength trivially all commute with one
another.
\begin{equation}
[{G}_{\alpha'\beta'}(x'),{G}_{\mu'\nu'}(x')] = i f^{abc} G^a_{\alpha'\beta'}
G^b_{\mu'\nu'} T^c \propto f^{abc}E^a E^b T^c = 0.
\end{equation}
Because the field strengths commute in one frame,
they must commute in all frames.
\begin{equation}
[{G}_{\alpha\beta}(x'),{G}_{\mu\nu}(x')] = \Lambda^{\alpha'}_\alpha
\Lambda^{\beta'}_\beta \Lambda^{\mu'}_\mu \Lambda^{\nu'}_\nu
[{G}_{\alpha'\beta'}(x'),{G}_{\mu'\nu'}(x')] = 0.
\end{equation}
Thus we may perform a gauge transformation upon ${G^a}_{\mu\nu} T^a$ to
rotate it into an abelian subalgebra of $SU(3)$.  That is, there exists a
gauge transformation $\Omega(x)$ such that the field strength can be
rotated to lie in, say, the ``three'' direction of $SU(3)$, $\Omega(x)
{G^a}_{\mu\nu}(x) T^a \Omega^{-1}(x) = G^3_{\mu\nu}(x) T^3$, or that the
field strength is purely in a $U(1)$ subalgebra of $SU(3)$.  Choosing the
field strength to lie in the ``three'' direction of $SU(3)$ partially fixes
the gauge, leaving a residual $U(1)$ gauge freedom.  We then make the
identification of the ``electromagnetic'' field strength as $F_{\mu\nu}
\equiv G^3_{\mu\nu}$.

\section{The effective Hamiltonian}
\label{effh}

We begin this section by establishing a relativistically valid four-vector
potential in the laboratory frame that yields ${\bf B}' = {\bf 0}$ in the
instantaneous rest frame of the quark.  While the four-vector potential is
not a gauge invariant object, the full wave equation for the moving quark
is gauge invariant provided that we perform a gauge transformation on the
quark wave function as well as on the four-vector potential.  In the
instantaneous rest frame the ${\bf E}'$ and ${\bf B}'$ fields at the quark
are assumed to be those given in the Buchm\"uller picture,
\begin{eqnarray}
{\bf B}' & = & {\bf 0} ,
\label{comb}\\
{\bf E}' & = & - a \hat{\bf r}' .
\label{come}
\end{eqnarray}
It is important to note that while the quark is acted upon by the
chromodynamic fields, it is a source of the chromodynamic fields as well.
The vector potential depends on the position and motion of the quark as
well as upon the field point ${\bf r}$
\begin{equation}
A_\mu = A_\mu({\bf r}, t,{\bf r}_Q,{\bf p}_Q).
\label{genA}
\end{equation}

The curl of the vector potential ${\bf A}$ with respect to the field point
is the magnetic field
\begin{equation}
{\bf B}({\bf r}, t,{\bf r}_Q,{\bf p}_Q) =
\nabla_{{\bf r}}\times{\bf A}({\bf r}, t,{\bf r}_Q,{\bf p}_Q).
\label{genB}
\end{equation}
It is not the general magnetic field (\ref{genB}) that appears in a wave
equation, but rather the operator expression
\begin{equation}
{\bf B}_{\rm eff}({\bf r}_Q,t,{\bf p}_Q) =
i({\bf p}_Q\times{\bf A}({\bf r}_Q,t,{\bf r}_Q,{\bf p}_Q) + {\bf A}({\bf
r}_Q,t,{\bf r}_Q,{\bf p}_Q
) \times{\bf p}_Q)= \nabla_{{\bf r}_Q}\times{\bf A}({\bf r}_Q,t,{\bf
r}_Q,{\bf p}_Q),
\label{Bsal}
\end{equation}
which we assume to be the magnetic field acting on the quark,
\begin{equation}
{\bf B}({\bf r}_Q, t,{\bf r}_Q,{\bf p}_Q) =
{\bf v}_Q\times{\bf E}({\bf r}_Q) .
\label{Bend}
\end{equation}

We find the simplest such potential yielding this magnetic field by
applying a Lorentz transformation to the co-moving potentials\footnote{ It
is straightforward to include an additive constant in the expression for
$A_0'$ which we neglect for simplicity.}
\begin{eqnarray}
A_0' &=& ar',\\
{\bf A}' &=& {\bf 0},
\label{Aprime}
\end{eqnarray}
and using the relation between momentum and quark velocity,
\begin{equation}
{\bf p}_Q = \gamma_Q m {\bf v}_Q + {\bf A},
\label{canmom}
\end{equation}
to obtain the potentials in the lab frame. The only quantities that enter
into the dynamical equations are functions of the quark position and
momentum so we may drop the subscripts $Q$ on quantities that refer to the
quark from now on as there is no more possibility of confusion.
\begin{eqnarray}
A_0({\bf r}',{\bf p}) &=& a\gamma r'
                   = a r' \sqrt{[1 + ({\bf p}/(m + ar'))^2]},
\label{Alab0}\\
{\bf A}({\bf r}',{\bf p}) &=& a\gamma {\bf v} r'
                   = {ar' \over m + ar'}{\bf p} .
\label{Alab}
\end{eqnarray}
We assume that ${\bf E}' \propto \hat{\bf r}'$ and ${\bf E} \propto \hat{\bf
r}$ so we must use the Lorentz transformed radius in the lab frame
\begin{equation}
r' = {r\over \gamma^{\phantom{2}}_\parallel} =
{r\sqrt{1 - ({\bf v}\cdot\hat{\bf r})^2}}.
\end{equation}
To lowest relativistic order, this correction is only important in $A_0$.

Buchm\"uller's argument predicts the Thomas spin-orbit long range
interaction, but it is not a complete dynamical theory.  In particular, an
effective Hamiltonian is needed to solve for energy levels, transition
rates, etc. To proceed in this direction we promote the lab frame
four-vector potential $A_{\mu}(r)$ ({\it i.e.\/}~the four-vector potential
at the quark position $r$) given in (\ref{Alab0}) and (\ref{Alab}), to a
suitably symmetrized operator, and use it with the Salpeter
equation\footnote{Use of the Salpeter equation avoids possible
inconsistencies of the Dirac equation due to the Klein paradox.  For
purposes of deriving a non-relativistic reduction either equation is
acceptable because both equations have the same nonrelativistic
reduction \cite{ovw}.}  \cite{salpeter,long} in the heavy-light limit
\cite{long},
\begin{equation}
\Lambda_{+}[{\mbox{\boldmath$\alpha$}} \cdot ({\bf p} - {\bf A} (r)) + \beta m
+ A_{0}(r){\openone} - E{\openone}]\Lambda_{+} \Psi = 0 .
\label{salp}
\end{equation}
As before, we have absorbed the charge into the potential.  In the above
equation (which is also known as the ``no-pair'' equation \cite{sucher}) we
have
\begin{eqnarray}
\Lambda_{\pm} &=& \frac{E_{0}\pm H_{0}}{2 E_{0}} ,
\label{lam}\\
H_{0} &=& {\mbox{\boldmath$\alpha$}} \cdot {\bf p} + \beta m , \\
E_{0} &=& \sqrt{p^{2}+m^{2}} .
\label{e0}
\end{eqnarray}
The angular momentum is given by
\begin{equation}
{\bf J} = {\bf r} \times {\bf p} + \frac{1}{2}\mbox{\boldmath$\sigma$} ,
\label{angmom}
\end{equation}
where the canonical momentum is the sum of the mechanical quark momentum
and the field momentum ${\bf A}$, as given in eq.~(\ref{canmom}).

If we define
\begin{equation}
\mbox{$\Bbb A$} \equiv
A_{0}\mbox{\openone}  - {\mbox{\boldmath$\alpha$}} \cdot {\bf A}  ,
\label{ii}
\end{equation}
where ${\openone}$ is the unit matrix, we can put (\ref{salp}) in normal
Salpeter form (with $E$ denoting the light quark energy)
\begin{equation}
(H_{0} + \Lambda_{+} \mbox{$\Bbb A$} \Lambda_{+} - E)\Psi = 0.
\label{salp2}
\end{equation}
Since the wave function already has the positive energy projected, it
follows that
\begin{equation}
\Lambda_{-} \Psi = \Lambda_{-} \left( \begin{array}{c}
f({\bf r}) \\
g({\bf r}) \end{array} \right) = 0,
\end{equation}
and hence
\begin{equation}
g({\bf r} ) = \frac{\mbox{\boldmath$\sigma$} \cdot {\bf p} }{E_{0}+m}
f({\bf r} ).
\label{gr}
\end{equation}
By using (\ref{gr}) in (\ref{salp2}) we obtain the two component form of
Salpeter equation
\begin{equation}
H f({\bf r} ) = E f({\bf r} ),
\end{equation}
where
\begin{equation}
H = E_{0} + (\Lambda_{+}\mbox{$\Bbb A$} \Lambda_{+})_{11} +
(\Lambda_{+}\mbox{$\Bbb A$} \Lambda_{+})_{12} \frac{\mbox{\boldmath$\sigma$}
\cdot {\bf p} }{E_{0}+m} .
\end{equation}
After a few lines of algebra, using (\ref{lam})-(\ref{e0}) and (\ref{ii}),
we obtain the effective Hamiltonian in the form
\begin{equation}
H = E_{0} + H_{E} + H_{M} ,
\label{effham}
\end{equation}
where
\begin{eqnarray}
H_{E} &=& \frac{1}{2 E_{0}}\left[ (E_{0}+m)A_{0} +
(\mbox{\boldmath$\sigma$} \cdot {\bf p} ) A_{0} (\mbox{\boldmath$\sigma$} \cdot
{\bf p} )\frac{1}{E_{0} + m}\right] , \label{effhamE}\\
\label{effhamM}
H_{M} &=& -\frac{1}{2 E_{0}}\left[(E_{0}+m)(\mbox{\boldmath$\sigma$} \cdot
{\bf A} ) (\mbox{\boldmath$\sigma$}
\cdot {\bf p} )\frac{1}{E_{0}+m} + (\mbox{\boldmath$\sigma$} \cdot {\bf p} )
(\mbox{\boldmath$\sigma$} \cdot {\bf A} ) \right] .
\end{eqnarray}
No gauge condition has been imposed upon the ($U(1)$) vector potential so the
effective Hamiltonian given in (\ref{effham})-(\ref{effhamM}) is valid for
any gauge.  The eigenstates of this Hamiltonian in two different gauges
will be related to each other by ($U(1)$) gauge transformations in the usual
way.

\section{Relativistic corrections}
\label{corrections}

To make contact with previous work we consider the case where the fermion
has a large mass $m$ and moves slowly.  We wish to find an effective action
that can be used perturbatively to compute relativistic corrections to the
non-relativistic reduction as an expansion in the small quantities $p/m$
and $a r/m$.

The expansion in $p/m$ and $a r/m$ is facilitated by a few commutator
identities.  Given that
\begin{equation}
[p^{2},F(r)] = -\nabla^2F(r) - 2 \frac{dF}{d r}\frac{\partial}{\partial r},
\end{equation}
and since the expectation values satisfy
\begin{equation}
2 \langle \frac{d F}{d r}  \frac{\partial }{\partial r} \rangle
= -\langle \nabla^{2} F\rangle ,
\end{equation}
we find
\begin{equation}
\langle [p^{2},F(r)] \rangle = 0 .
\end{equation}
Hence, to lowest order in
$p^{2}/m^{2}$ ($E_{0}^{\pm 1} \simeq m^{\pm 1}
(1 \pm p^{2}/2 m^{2}\ldots)$),
\begin{equation}
\langle [E_{0}^{\pm 1}, F(r)]\rangle \simeq 0 .
\end{equation}
It is this relation that guarantees that the Dirac equation and the
Salpeter equation have the same non-relativistic reduction.

With these aids we can immediately reduce the terms of the effective
Hamiltonian (\ref{effham}),
\begin{eqnarray}
E_{0} &\simeq& m + \frac{p^{2}}{2 m} - \frac{p^{4}}{8 m^{3}} ,\\
H &\simeq&
E_{0} + A_{0} + \frac{\nabla^{2}A_{0}}{8 m^{2}} -\frac{1}{2m}({\bf p}\cdot
{\bf A} + {\bf A}\cdot{\bf p}) + \frac{1}{2 m^{2}r} \frac{d A_{0}}{d r} {\bf s}
\cdot {\bf L} - \frac{1}{m} {\bf s} \cdot \nabla \times {\bf A} .
\label{effh2}
\end{eqnarray}

The potentials we found are functions of quark position and momenta, which
is a situation not often encountered in electrodynamics.  To construct a
quantum theory with Hermitian potentials, we symmetrize the products of
position and momentum.  Ideally we would use Weyl ordering to make
connection with the midpoint prescription for evaluating a path
integral. Using (\ref{Alab0}) and (\ref{Alab}), we find
\begin{eqnarray}
\nabla \times {\bf A}(r) &\simeq& \frac{a {\bf L} }{mr}\ ,
\label{redcurla}\\
A_{0}(r) &\simeq& a r -{a\over 4m}({\bf r}\cdot{\bf p} + {\bf p}\cdot{\bf
r}) + {a\over 8 m^2 }({1\over r} r_ir_jp_ip_j +  p_ip_jr_ir_j{1\over r}) + \\
& & + \frac{a}{4m^2}({\bf p}^2 r + r {\bf p}^2)\ ,
\label{reda0}\\
{\bf A} &\simeq & \frac{a}{2m}({\bf p}r + r{\bf p})\ ,\\
\nabla ^{2} A_{0}(r) & \simeq & \frac{2 a}{r}\ .
\label{redai}
\end{eqnarray}
The second and third terms in $A_0$ result from the
$1/\gamma^{\phantom{2}}_\parallel$ correction to the radius.

With these expressions the reduction (\ref{effh2}) becomes
\begin{eqnarray}
H &\simeq& E_{0} + ar - \frac{3 a}{2 m^{2} r} + \frac{a r {p}_r^2}{2m^2} -
\frac{a r {\bf p}^2}{2m^2} - \frac{a}{2 m^{2} r} {\bf s} \cdot {\bf L},\\
&\simeq& E_{0} + ar - \frac{3 a}{2 m^{2} r} -
\frac{a L^2}{2m^2r} - \frac{a}{2 m^{2} r} {\bf s} \cdot {\bf L} .
\label{redham}
\end{eqnarray}

For contrast, we compare these results to those obtained in scalar and
time-component vector confinement and in QCD.

If the confinement is pure time component vector we substitute
\begin{equation}
A_0 = ar,\quad {\bf A} = 0,
\end{equation}
into (\ref{effh2}) to obtain
\begin{equation}
H_{\rm vector} \simeq E_{0} + ar + \frac{a}{4 m^{2} r}
+ \frac{a}{2 m^{2} r} {\bf s} \cdot {\bf L} .
\label{vecredham}
\end{equation}

In the case of scalar confinement we take
\begin{equation}
\mbox{$\Bbb A$} = \beta a r.
\end{equation}
A similar reduction gives
\begin{equation}
H_{\rm scalar} \simeq E_{0} + ar - \frac{a}{4 m^{2} r} -
\frac{a r {p}_r^2}{4m^2} -
\frac{a L^2}{4m^2r} - \frac{a}{2 m^{2} r} {\bf s} \cdot {\bf L} .
\label{scalarredham}
\end{equation}

Finally, the low quark velocity expansion of the Wilson minimal area law
provides the effective Hamiltonian implied by QCD \cite{prosperi},
\begin{equation}
H_{\rm QCD} \simeq E_{0} + ar + \frac{a}{36 m^{2} r} -
\frac{a L^2}{6 m^2r} - \frac{a}{2 m^{2} r} {\bf s} \cdot {\bf L} .
\label{qcdredham}
\end{equation}

We observe that our model, scalar confinement, and low quark velocity
expansion of the Wilson loop model \cite{prosperi} all have the pure Thomas
spin-orbit interaction.  The Darwin terms differ but are sensitive to the
ordering prescription applied to symmetrize the operators.  We note that
even in the case where numerical coefficients do not agree, our model has
coefficients of the same sign as both the scalar confinement and Wilson
loop results.

\section{Regge Behavior}
\label{regge}

The slope of the leading Regge trajectory can be calculated classically in
the limit of highly excited circular orbits as
\begin{equation}
\alpha' \equiv {L \over E^2},
\end{equation}
where $E$ is  the excitation energy and  $L$ is the orbital angular
momentum.  For such orbits spin and quark mass are negligible and, by
(\ref{Alab0}) and (\ref{Alab}), we have in that limit
\begin{eqnarray}
A_0 & \longrightarrow & \sqrt{a^2 r^2 + {\bf p}^2},\label{massless0}\\
{\bf A} & \longrightarrow & {\bf p}.
\label{massless}
\end{eqnarray}
In this limit, the  Salpeter equation (\ref{salp}) requires that
\begin{equation}
E  =  A_0.
\end{equation}
Using (\ref{massless0}), we find
\begin{equation}
E^2 = (ar')^2 + {\bf p}^2 \simeq (ar')^2 + ({L\over r'})^2.
\end{equation}
The leading trajectory is given by the lowest energy for a fixed angular
momentum \hbox{${\partial E \over \partial r'}\Big |_L = 0$}, so
\begin{equation}
L= ar'.
\end{equation}
With the circular orbit condition, $p = L/r'$, this gives
\begin{equation}
  E^2 = 2 a L,
\end{equation}
and the Regge slope becomes
\begin{equation}
\alpha' = {L \over E^2} = {1 \over 2 a}.
\end{equation}

We note that this is also the result given by the Dirac equation with
scalar confinement \cite{ovw} in a heavy-light meson and that time
component vector confinement gives a Regge slope of
\begin{equation}
\alpha'_{\rm vector} = {1\over 4 a} .
\end{equation}

As demonstrated analytically in \cite{ovw}, scalar confinement with the
Salpeter equation does not give the desired linear trajectories.

\section{Conclusion}
\label{con}

Our starting point here has been Buchm\"uller's physical picture relating
the static chromoelectric flux tube and the resulting pure Thomas
spin-orbit confinement interaction in a meson, though in this paper we make
use only of the result that the chromomagnetic field vanishes at the quark
position.  When this assumption is expressed in the laboratory frame, its
full content becomes more apparent. Once the lab frame four vector
potential is found by Lorentz transforming a purely electric potential in
the instantaneous quark rest frame to the laboratory frame, a complete
relativistic effective Hamiltonian can be directly computed. Not only is
the expected Thomas spin-orbit term found, but all other relativistic
corrections can be evaluated. 

The principal virtue of the model presented here is that it has very few
assumptions and further that these assumptions are strongly motivated by
QCD.  We assume that the only field acting on the quark is a constant
radial electric field in its rest frame and that the magnetic field at the
quark vanishes in its rest frame.  From these assumptions follow the Thomas
type spin-orbit interaction that is strongly favored by lattice simulations
and phenomenological considerations, as well as linear Regge trajectories.
An additional virtue of the model is its simplicity.  It is essentially a
potential model reducing to the solution of a radial differential equation
in the general case.

For reason of simplicity and clarity of exposition, we have considered one
quark to be fixed.  Just as for all potential models, our results carry
over directly to the two body case with suitable introduction of center of
momentum and relative coordinates.  For a realistic meson we must also
include a short range vector (Coulomb) interaction. In the heavy-light
limit this short range term will result in a spin-orbit interaction
opposite in sign to the Thomas term.

\vskip 1cm
\begin{center}
ACKNOWLEDGMENTS
\end{center}
One of us (MGO) would like to thank Prof.~C. Goebel for helpful
conversations.  This work was supported in part by the U.S. Department of
Energy under Contract Nos.  DE-FG02-95ER40896 and DE-AC05-84ER40150, the
National Science Foundation under Grant No. HRD9154080, and in part by the
University of Wisconsin Research Committee with funds granted by the
Wisconsin Alumni Research Foundation.

\end{document}